\documentclass[prd,tightenlines,showpacs,nofootinbib,eqsecnum,amsfonts,amssymb,amsmath,twocolumn,widetext]{revtex4}
\usepackage{graphicx}
\usepackage{amsmath}
\usepackage{amsfonts}   
\usepackage{amssymb}  
\usepackage{mathrsfs}
\usepackage{epsfig,bbm}
\usepackage{bm}
\usepackage{color}
\usepackage{dsfont}
\begin{document}
\newcommand{\average}[1]{\langle{#1}\rangle_{{\cal D}}}
\newcommand{\dd}{{\rm d}}
\newcommand{\etal}{{\it et al.}}
\title{Late time anisotropy as an imprint of cosmological backreaction}

\author{Giovanni Marozzi}
\email{giovanni.marozzi@college-de-france.fr}
\affiliation{
		Coll\`ege de France, 11 Place M. Berthelot, 75005 Paris, France,
             }

\author{Jean-Philippe Uzan}
\email{uzan@iap.fr}
 \affiliation{
             Institut d'Astrophysique de Paris,
             Universit\'e Pierre~\&~Marie Curie - Paris VI,
             CNRS-UMR 7095, 98 bis, Bd Arago, 75014 Paris, France.}

\begin{abstract}
Backreaction effects of the large scale structure on the background dynamics have been claimed to lead to a renormalization of the background dynamics that may account for the late time acceleration of the cosmic expansion. This article emphasizes that generically the averaged flow is locally anisotropic, a property that can be related to observation. Focusing on perturbation theory, the spatially averaged shear, that characterizes the anisotropy of the flow, is computed.  It is shown that this shear arising from backreaction differs from a homogeneous shear: its time evolution is different and its amplitude is completely determined by the cosmological parameters and the matter power spectrum. It ranges within (2-37)\% at a redshift of order 0.5 so that the isotropy of the Hubble flow may allow to constrain the backreaction approach to dark energy.
\end{abstract}
\date{21 June 2012}
\maketitle

\section{Introduction}

In the standard cosmological framework~\cite{pubook}, one assumes that on large scales, the universe is well described by a spatially homogeneous and isotropic spacetime, at least at the background level, so that its dynamics is obtained from the Einstein equations for a Friedmann-Lema\^{\i}tre (FL) metric. Structure formation is then described using perturbation theory, which has proven to be successful to understand the existing observations from cosmic microwave background anisotropies to the growth rate of the large scale structure.

While most of the observations are compatible with the assumption of a spatially homogeneous and isotropic universe on large scales, they exhibit a clumpy distribution of matter on small scales and at late time. This is at the heart of a lively debate concerning the magnitude of the backreaction of the large scale structure on the background dynamics. While its magnitude depends on the averaging procedure~\cite{Buchert,Zalaletdinov} (see also Ref.~\cite{PS}) and on the actual small scale geometry of the universe, it has been mostly estimated using perturbation theory at linear and second order either in synchronous gauge~\cite{LS,rasanen1,Kolb:2004am} or in Newtonian gauge~\cite{Kolb:2004am,rasanen2,behrend,brown,kasai,CAN} (see e.g. Ref.~\cite{bura} for reviews). One important conclusion is the existence of ultraviolet divergences (see, for example, Refs.~\cite{Kolblast,CU}) when particular observables are calculated, with no convincing regularization schemes proposed so far. This lets open the question of the magnitude of the backreaction and its ability to explain the late time acceleration of the cosmic expansion.
This is a rather controversial subject and, according to some authors \cite{rasanen1,rasanen2,AutPro} the backreaction of present inhomogeneities might explain, by itself, cosmic acceleration while,  according to others \cite{AutAga}, the effect of inhomogeneities is totally negligible.

This article lies on a very simple, and almost trivial, remark: the spatially averaged flow has no reason to be isotropic. While obvious, this fact has been hidden in the generally adopted backreaction procedure~\cite{Buchert}, in which the averaged dynamics is presented in a form mimicking the Friedmann dynamics so that attention has been focused on the volume averaged expansion factor and on the backreaction terms (see below for definitions). Late time growth of a spatial anisotropy may thus be a specific signature of backreaction. The goal of this work is thus to estimate the level of anisotropy of the averaged flow and address the following questions: (1) Is the shear, which is related to the anisotropy of the Hubble flow (and thus observable), divergent in the ultraviolet? (2) How is the shear magnitude related to the backreaction magnitude? (3) Can a bound on the shear on cosmological scales allow us to set a bound on the backreaction?

To that purpose, we start by recalling the averaging procedure in \S~\ref{II} and propose to split the kinematical backreaction term into a term describing the anisotropy of the flow and a genuine backreaction term. Under that form the averaged dynamics of any spatially homogeneous (but not necessarily isotropic) flow on a surface of homogeneity remains unchanged and the genuine backreaction term strictly vanishes. Using perturbation theory, the shear is computed at lowest order in perturbation theory in \S~\ref{III} and then computed explicitly in \S~\ref{V} after having discussed the choice of the matter power spectrum in \S~\ref{IV}. We have actually used two power spectra, one motivated by observations (with a simple analytical form allowing for an exact integration) and another motivated by theory, relying on the initial conditions from inflation and the transfer function of the standard $\Lambda$CDM model, but which requires numerical integration. As we shall show the scalar shear is divergent in the ultraviolet, which raises several questions that are summarized in \S~\ref{VI}. Appendix A summarizes the main results of linear perturbation theory around a spatially Euclidean homogeneous background spacetime. A detailed description of how to implement the shear average is given in Appendix B, and the normalization of the power spectra is described in Appendix C.


\section{Averaged cosmological dynamics}\label{II}

\subsection{Spacetime foliation}

The driving idea, as first proposed in Ref.~\cite{Buchert}, has been to rely on the 1+3 splitting of the universe~\cite{1plus3} associated with a general reference timelike congruence $n^\mu$ that defines a foliation of spacetime. The choice of this congruence is often referred to as a choice of observers. The 3-dimensional spacelike hypersurfaces normal to $n^\mu$ can then be defined by the equation $S({\bm x}, t)-S_0=0$, with $S({\bm x}, t)$ a scalar field and $S_0$ a constant. Then
\begin{equation}\label{DefnA}
n_{\mu} \equiv -  \frac{\partial_{\mu} S}{ (-
 \partial_{\mu} S \partial_{\nu} S ~ g^{\mu\nu}) ^{1/2}},
\end{equation} 
which is normalized as $n_{\mu} n^{\mu} = -1$. This allows us to define $h_{\mu\nu}$, the projector on these hypersurfaces, as
\begin{equation}
  h_{\mu\nu} = g_{\mu\nu} + n_{\mu}n_{\nu}
\end{equation}
which satisfies by construction $h_{\mu\rho}h^{\rho}_{\nu} = h_{\mu\nu}$ and $h_{\mu\nu}n^{\mu} = 0$. One can then define 
the expansion $\Theta$, shear $\sigma_{\mu\nu}$ and vorticity $\omega_{\mu\nu}$ of the flow as
\begin{eqnarray}
   \Theta_{\mu\nu} &\equiv& h^\alpha_\mu h^\beta_\nu \nabla_\alpha n_\beta\\
                               & =&\frac{1}{3} h_{\mu\nu}\Theta+\sigma_{\mu\nu}+\omega_{\mu\nu} .
\end{eqnarray}
They are explicitly given by
\begin{equation}\label{e.theta}
    \Theta \equiv \nabla_\mu n^\mu,
\end{equation}    
\begin{equation}
  \sigma_{\mu\nu}\equiv h^\alpha_\mu h^\beta_\nu \left[\nabla_{(\alpha} n_{\beta)}-
  \frac{1}{3} h_{\alpha\beta}
\nabla_\tau n^\tau \right],
\end{equation}
\begin{equation}
\omega_{\mu\nu}\equiv h^\alpha_\mu h^\beta_\nu \nabla_{[\alpha} n_{\beta]}\,.
\end{equation}
Indeed, the assumption of Eq. (\ref{DefnA}) implies that the vorticity strictly vanishes, $\omega_{\mu\nu}=0$. 


\subsection{Buchert's formalism in a nutshell}

In practice, perturbations grow significantly only during the matter-dominated era, so that one can restrict the analysis to dust-filled universes, eventually with a cosmological constant. In such a situation, and as it was assumed in the original work by Buchert~\cite{Buchert}, one can pick up a foliation defined by a geodesic congruence $n_\mu$ which accidentally coincides with the four-velocity $u_\mu$ of comoving observers, i.e. $n_\mu=u_\mu$.\footnote{In
general $n_\mu$, which defines a general reference flow, and $u_\mu$, which defines the four-velocity of the 
observers comoving with the matter, may be different (see Ref.~\cite{GMV2} for details).}

The shear tensor can be expressed as
\begin{equation}
 \sigma_{\mu\nu}=\Theta_{\mu\nu}-\frac{1}{3} h_{\mu\nu}\Theta
\end{equation}
and it follows that the scalar shear takes the form
\begin{equation}
  \sigma^2\equiv\frac{1}{2} \sigma^\mu_\nu \sigma^\nu_\mu=
\frac{1}{2} \left(\Theta^\mu_\nu \Theta^\nu_\mu-\frac{1}{3}
\Theta^2\right)\, .
\label{sigma2}
\end{equation}

The spatial average of any scalar quantity $A$ on a domain ${\cal D}$ is then usually defined as
~\cite{Buchert}
\begin{equation}
 \average{A(t,{\bm x})} = \frac{1}{V_{\cal D}}\int_{\cal D} \sqrt{|h|}\, A(t,{\bm x})\dd^3{\bm x} 
\label{averagepb}
\end{equation}
where $V_{\cal D}$ is the volume of the domain, defined by the requirement that $\average{1}=1$, 
and $h$ is the determinant of the induced metric $h_{\mu\nu}$. 
Such a spatial average is associated with the general reference timelike congruence $n^\mu$
of Eq.~(\ref{DefnA}) if and only if the average is performed in the gauge where $S({\bm x}, t)$ is 
homogeneous (see Refs.~\cite{GMV1,Marozzi} and Appendix B). 

The average~(\ref{averagepb}) does not commute with the time derivative and gives origin to the so-called Buchert-Ehlers commutation
rule \cite{BE}
\begin{equation}
 \partial_t  \average{A(t,{\bm x})} -  \average{\partial_t A(t,{\bm x})} =  \average{\Theta A}- \average{\Theta} \average{A}
 \label{BEequation}
\end{equation}
where $\average{\Theta}$ is the spatial average of the expansion $\Theta$ defined in Eq.~(\ref{e.theta}). 
We can then define an averaged scale factor $a_{\cal D}$, related to the volume $V_{\cal D}$ of the integration domain (normalized by a reference volume scale $V_{{\cal D}_0}$), by $a_{\cal D} \equiv (V_{\cal D}/V_{{\cal D}_0})^{1/3}$. Considering Eq.~(\ref{BEequation}), we then have the useful relation
\begin{equation}
 \average{\Theta} \equiv 3\frac{\dot a_{\cal D}}{a_{\cal D}}\,,
\end{equation}
where it is important to emphasize that this scale factor describes the averaged dynamics of the flow and cannot be identified with a metric component of a FL spacetime. 

The bottom line of the averaging procedure proposed in Ref.~\cite{Buchert} is to allow for the spatial averaging of scalar quantities and to derive an averaged dynamical flow by averaging the generalized Friedmann and Raychaudhuri equations to get 
\begin{eqnarray}
\!\!\!\!\!\!\!\!\!\! && \left(\frac{\dot a_{\cal D}}{a_{\cal D}}\right)^2 = \frac{8\pi G}{3}\average{\rho} + \frac{\Lambda}{3} -\frac{1}{6}\left({\cal Q}_{\cal D} +\average{\cal R} \right)\label{e.backreac1}\\
\!\!\!\!\!\!\!\!\!\! && \frac{\ddot a_{\cal D}}{a_{\cal D}} = -\frac{4\pi G}{3}\average{\rho} + \frac{\Lambda}{3} -\frac{{\cal Q}_{\cal D}}{3}\label{e.backreac2}
\end{eqnarray}
where one has defined the kinematical backreaction term as
\begin{equation}\label{e.QD}
 {\cal Q}_{\cal D} \equiv \frac{2}{3}\left(\average{\Theta^2}-\average{\Theta}^2 \right) - 2\average{\sigma^2}
\end{equation}
and $\average{\cal R}$ refers to the spatial average of the curvature of the 3-dimensional metric induced on the hypersurfaces.
Since the average of the continuity equation, $\dot\rho+\Theta\rho=0$, leads to
\begin{equation}
 \partial_t\average{\rho}+\average{\Theta}\average{\rho}=0,
\end{equation}
one easily concludes that $\average{\rho}\propto a_{\cal D}^{-3}$.

This formalism has been extended to allow for arbitrary coordinate systems~\cite{CAN,Larena:2009md} and any choice of gauge and slicing~\cite{GMV1,GMV2}.


\subsection{Rewriting of the averaged equations}

In the form~(\ref{e.backreac1}--\ref{e.backreac2}), the dynamics of the averaged scale factor mimics the Friedmann equations for a spatially homogeneous and isotropic FL spacetime. The dynamics of the averaged flow depends strongly on the backreaction terms ${\cal Q}_{\cal D}$ and $\average{{\cal R}}$. These terms are related each other by the integrability condition 
\begin{equation}
 \frac{1}{a_{\cal D}^6}\partial_t\left({\cal Q}_{\cal D}a_{\cal D}^6\right) +  \frac{1}{a_{\cal D}^2}\partial_t\left(\average{{\cal R}}a_{\cal D}^2\right) =0
\end{equation} 
but there remains a freedom in the determination of ${\cal Q}_{\cal D}$. 
To this purpose, the value of ${\cal Q}_{\cal D}$ has been computed using perturbation theories in many studies (see, for example, Refs.\cite{LS,rasanen1,Kolb:2004am,rasanen2,behrend,brown,kasai,CAN,Kolblast,CU}), and
many alternative models have been proposed (see, for example, Ref.~\cite{RB}).

The form~(\ref{e.backreac1}--\ref{e.backreac2}) and the splitting of the backreaction terms is highly suggestive since they match a Friedmann form. It is however intuitively clear that the average of any spatially homogeneous flow should let this flow unchanged. In particular, we would like the kinematical backreaction term to strictly vanish when averaging any spatially homogeneous flow. To that purpose, we rewrite the system~(\ref{e.backreac1}--\ref{e.backreac2}) in the strictly equivalent form
\begin{eqnarray}
& & \!\!\!\!\!\!\!\!\!\!\!\!\!\!\!\!\!\!\!\!\left(\frac{\dot a_{\cal D}}{a_{\cal D}}\right)^2\!\!=\!\frac{8\pi G}{3}\average{\rho}\! + \!\frac{\Lambda}{3}\! +\! \frac{\average{\sigma^2}}{3}
 \!  
 -\frac{1}{6}\!\left(\!\tilde{\cal Q}_{\cal D} +\average{\cal R}\!\right)
 \label{e.backreac1bis}\\
& &  \!\!\!\!\!\!\!\!\!\!\!\!\!\!\!\!\!\!   \frac{\ddot a_{\cal D}}{a_{\cal D}}= -\frac{4\pi G}{3}\average{\rho} + \frac{\Lambda}{3} +\frac{2}{3}\average{\sigma^2}
 - \frac{\tilde{\cal Q}_{\cal D}}{3},\label{e.backreac2bis}
\end{eqnarray}
defining a new kinematical backreaction term as
\begin{equation}\label{e.QD}
 \tilde{\cal Q}_{\cal D} \equiv \frac{2}{3}\left(\average{\Theta^2}-\average{\Theta}^2 \right).
\end{equation}
Under this form, the averaged flow is described by a set of equations that mimics the flow of comoving observers in a spatially homegenous universe, but not necessarily isotropic spacetime. It follows that the backreaction term $\tilde{\cal Q}_{\cal D}$ is generated only by inhomogeneities and is thus related to deviations from local homogeneity.

Indeed, it is clear that for a FL universe, $\average{\sigma^2}=0$ while when averaging any Bianchi universe, $\average{\sigma^2}=\sigma^2$ and $\tilde{\cal Q}_{\cal D}=0$.
In both cases the set of equations (\ref{e.backreac1bis}--\ref{e.backreac2bis}) 
has the homogeneous form, while the standard kinematical backreaction ${\cal Q}_{\cal D}$ 
is different from zero for the later case despite no genuine backreaction. 
With such a rewriting it is possible to compare the backreaction terms induced by the perturbations on a background flow with either FL or Bianchi symmetry~\cite{tocome} but more important to compare the magnitude of the anisotropy of the averaged flow with the true backreaction term.

The averaging of spatially anisotropic cosmological models has been considered in Ref. \cite{BT}.
Our definition of $\tilde{\cal Q}_{\cal D}$, and the split of the kinematical backreaction term in 
$\tilde{\cal Q}_{\cal D}$ and $\average{\sigma^2}$, differs from their proposal.
We emphasize that their definition relies on the use of $\average{\sigma}^2$, where $\sigma$ itself is 
not well-defined, contrary to $\sigma^2$.
For instance, considering a FL background, the lowest order for $\sigma^2$ will be quadratic in 
first order perturbations (see Appendix A and next section) which are stochastic fields and one 
needs to define properly the square root of these stochastic fields before ensemble average.
Indeed, starting with a non-isotropic background, one can define $\sigma$ as $\sqrt{(\sigma^2)^{(0)}}$
at lowest order and then perform a perturbative expansion.
Given this, we prefer to adopt a more straightforward and pedestrian rewriting of the standard 
equation.


\section{Quantifying the deviation from isotropy}\label{III}

As explained above, the goal of this study is to quantify the anisotropy of the averaged flow. For that purpose, we start from a FL universe with perturbations in order to compute $\average{\sigma^2}$. In such an approach, one needs to be reminded that the perturbations are stochastic fields, usually with Gaussian initial conditions. It follows that the spatially averaged quantities are also stochastic quantities.

Taking $X$ as a function of the perturbations and $\average{X}$ as its average on a spatial domain, then, from a theoretical point of view, we only have access to the distribution of $\average{X}$, that is to $\overline{\average{X}}$, which is the ensemble average of $\average{X}$. In this section, we compute the lowest contribution to the shear in perturbation theory. The perturbation equations at linear order are summarized in \S~\ref{III.b} and the shear is explicitly computed in \S~\ref{III.c}.


\subsection{Background dynamics}

For simplicity, we consider a spatially Euclidean Friedmann-Lema\^{\i}tre universe so that the late time dynamics is dictated only by the pressureless matter and the cosmological constant with density parameters
\begin{equation}
 \Omega_{\rm m0} = \frac{8\pi G\rho_{\rm m0}}{3H_0},\qquad
 \Omega_{\Lambda0} = \frac{\Lambda}{3H_0}
\end{equation}
that satisfy $\Omega_{\rm m0}+ \Omega_{\Lambda0}=1$. The Friedmann equation then takes the usual form
\begin{equation}
 E(z) \equiv \frac{H^2(z)}{H_0^2} =   \Omega_{\rm m0}(1+z)^3 + \Omega_{\Lambda0},
\end{equation}
the solution of which is
\begin{equation}
 a(t) \propto \sinh^{2/3}\left(\frac{3}{2}\sqrt{\Omega_{\Lambda0}} H_0t \right).
\end{equation}
The normalization to the Hubble constant today, $H_0$, implies that
\begin{equation}
  \sinh\left(\frac{3}{2}\sqrt{\Omega_{\Lambda0}} H_0t_0 \right) = \frac{\Omega_{\Lambda0}^{1/2}}{(1-\Omega_{\Lambda0})^{1/2}}\equiv \kappa_0^{3/2}
\end{equation}
so that the redshift is given by
\begin{equation}
 1+z =\frac{\kappa_0}{\sinh^{2/3}\left(\frac{3}{2}\sqrt{\Omega_{\Lambda0}} H_0t \right)}.
\end{equation}


\subsection{Perturbation equations}\label{III.b}

For our concern, it will be sufficient to assume that our Universe can be  described by a spatially homogeneous and isotropic FL spacetime with  perturbations.

We decide to fix the gauge freedom by explicitly choosing the gauge in which the scalar $S({\bm x},t)$ is homogeneous.  In this way we can use the simple definition of Eq.~(\ref{averagepb}) for the spatial average of a general scalar (see Appendix B). 
As already mentioned,
considering a $\Lambda$CDM model and the scalar $S({\bm x},t)$ which defines the reference hypersurface comoving with the matter, 
namely choosing the matter comoving gauge in which one imposes that $T^{i}_0=0$ (with $i=1,\ldots,3$), we choose observers which are also geodesic. 
We fix the remaining gauge freedom by imposing that $E=0$ and $\bar{E}_i=0$ (see Appendix A for the definitions). The metric thus takes the form
\begin{eqnarray}
\dd s^2 &=& a^2\left[-\dd\eta^2 + 2(\partial_i B + \overline B_i)\dd\eta\dd x^i \right.\nonumber\\
 &&\left.\qquad+\left(1+ 2 C \delta_{ij}+2 \overline{E}_{i j} \right]\dd x^i\dd x^j
\right]
\label{metric}
\end{eqnarray}
where $B$ and $C$ are scalars, $\overline{B}_i$ is a transverse vector 
($\partial_i \overline{B}^i=0$) and 
$\overline{E}_{ij}$ is a traceless and transverse tensor ($\partial_i \overline{E}^{ij}=0=\overline{E}_i^i$). The gauge invariant Bardeen potentials are then given by
\begin{eqnarray}
\Psi \equiv -C-\mathcal{H} B, \qquad
\Phi \equiv \mathcal{H} B+B'\,
\end{eqnarray}
and the density contrast and peculiar velocity by
\begin{eqnarray}
\delta^C=\delta, \quad V=v=-B \,.
\label{RelationVandB}
\end{eqnarray}

The perturbation equations reduce to the four Einstein equations and the Euler equation (again, we restrict ourselves to a dust plus cosmological constant universe, where the sound velocity $c_s$ is zero)
\begin{eqnarray}
  & & \nabla^2 \Psi=\frac{1}{2 M_{\rm Pl}^2} a^2 \rho \delta^C \label{Eq1} \\
  & & \Psi=\Phi \label{Eq2} \\
  & & \Psi'+ \mathcal{H} \Psi=-\frac{1}{2 M_{\rm Pl}^2}a^2 \rho V \label{Eq3}\\
  & & \Psi''+3 \mathcal{H} \Psi'+[2 \mathcal{H}'+\mathcal{H}^2] \Psi=0 \label{Eq4} \\
  & & V'+ \mathcal{H} V=-\Phi\,, \label{Eq5}
\end{eqnarray}
with $M_{\rm Pl}^{-2}=8\pi G$.


\subsection{Scalar shear}\label{III.c}

Following the definition (\ref{sigma2}) the scalar shear is 
by construction a second order quantity,
\begin{equation}
 (\sigma^2)^{(0)}= (\sigma^2)^{(1)}=0.
 \label{shear0and1MT}
\end{equation}
As a consequence, it can be specified to its lowest order using only first order perturbation theory. Considering the particular gauge choice specified in Eq.~(\ref{metric}), we obtain (see Appendix A for the expression in a general gauge)
\begin{eqnarray}\label{sigma2order2generalMT}
  2 a^2(\sigma^2)^{(2)} &=&
    B_{i, j} B^{i, j}-\frac{1}{3}(\partial^i B_i)^2 \nonumber \\ 
   & & -2 B_{i, j} {\bar{E}}^{i j\,'} + {\bar{E}}_{i j}^{'} {\bar{E}}^{i j\,'}
\end{eqnarray}
where $B_i\equiv\partial_i B+\bar{B}_i$ and we use the notation $X_{,i}\equiv \partial_i X$ for 
any field $X$. 

At first order in perturbations, vector and tensor modes are negligible compared to scalar modes so that we can safely neglect them. The average on a domain
${\cal D}$ is then obtained by inserting the expression~(\ref{sigma2order2generalMT}) in Eq.~(\ref{averagepb}),
\begin{equation}\label{sigma_starting_eq}
  \langle \sigma^2 \rangle_{\cal D}= \frac{1}{V_{\cal D}}\int_{\cal D} \frac{\dd^3{\bm x} }{2 a^2} \left[B_{,i j}
    B^{,i j}-\frac{1}{3}(\nabla^2 B)^2\right].
\end{equation}
Its ensemble average is easily obtained by shifting to Fourier space (see Appendix B for details) to get
\begin{eqnarray}\label{Res_EA}
  \overline{\langle \sigma^2 \rangle}_{\cal D}(\eta) &=& \frac{1}{3 a^2(\eta)}\int \frac{d^3{\bm k}}{(2 \pi)^{3}} k^4 |B_{\bm k}(\eta)|^2 
\end{eqnarray}
where, for any field $X$, $X_{\bm k}$ denotes its Fourier modes and $\overline{X_{\bm k}X_{\bm k'}}=\left\vert X_k\right\vert^2\delta^{(3)}({\bm k}+{\bm k}')$, because of statistical homogeneity and isotropy. The former integral can be evaluated by the use of Eqs.~(\ref{RelationVandB}), (\ref{Eq2}) and ~(\ref{Eq5}) so that the Fourier components of $B$ and $\Psi$ are related by 
\begin{equation} \label{B_Fourier}
  B_{\bm k}(\eta) =\frac{1}{a(\eta)} \int^\eta a(\eta') \Psi_{\bm k}(\eta')\dd\eta' . 
\end{equation}
As already concluded in many works~\cite{LS,CAN,CU} and shown here in the Appendix B, the ensemble average of a spatial average of any second order quantity on a domain ${\cal D}$ does not depend on the size and shape of the domain at lowest order in perturbation theory.

The anisotropy of the flow is characterized by the dimensional quantity 
\begin{equation}
\Sigma \equiv 3\frac{{\average{\sigma^2}}}{{(\average{\Theta})^2}}.
\end{equation}
In perturbation theory it is given by
\begin{equation}
 \Sigma = \frac{1}{3}\frac{{\average{(\sigma^2)^{(2)}}}}{H^2}\left(1 -\frac{2}{3}
 \frac{{\average{(\Theta^{(2)})}}}{H} + \ldots\right)
\end{equation}
so that at lowest order, it reduces to
\begin{equation}
 \Sigma = \frac{1}{3}\frac{{\average{(\sigma^2)^{(2)}}}}{H^2}.
 \end{equation}


\section{Choice of the power-spectrum}\label{IV}

To go further and compute the quantities derived in the previous section, we need to specify the power spectrum. 

We shall consider two power spectra. The first one relies on the inflationary prediction and the transfer function of the standard $\Lambda$CDM model~\cite{EH}. Our second choice is obtained from observation, and we decide to use the APM power spectrum~\cite{APM}. Even if it is not the most up-to-date observational power spectrum, it gives a good description and has a particular simple form that allows us to perform the computations analytically. The two choices are complementary and will allow us to get a good estimate of the order of magnitude of the averaged shear.

The spectrum $P_\delta (k, \eta)$ of the matter density contrast with Fourier components $\delta_{\bm k}(\eta)=\delta \rho_{{\rm m} \bm{k}}(\eta)/\rho_{\rm m}(\eta)$ is defined from the two-point correlation function
\begin{eqnarray}
  \overline{\delta_{\bm k}(\eta) \delta_{\bm k'}(\eta)} \equiv P_\delta(k,\eta) \delta^{(3)}(\bm{k}+\bm{k}')\,,
\end{eqnarray}
and we also introduce the notation
\begin{equation}
 {\cal P}_\delta \equiv \frac{k^3}{2\pi^2}P_\delta
\end{equation}
and use similar definitions for any field. 

The power-spectra of $\delta$ and $\Psi$ are indeed related since Eq. (\ref{Eq1}) implies that
\begin{equation}\label{ConnectionPowerSpectra}
  {\cal P}_\Psi =\frac{1}{2\pi^2} \left(\frac{1}{2 M_{\rm p}^2}\right)^2 \frac{a^4 \rho_{\rm m}^2}{k} P_\delta\,.
\end{equation}


\subsection{First choice: theory based}\label{subsec4a}

A first and natural choice is to use the power spectrum that is deduced from inflation and the transfer function computed in a $\Lambda$CDM model.
Following Ref.~\cite{CAN} the power spectrum takes the form
\begin{equation}\label{PowerPsiStandard}
  {\cal P}_{\Psi}=\left(\frac{3 \Delta_R}{5 g_{\infty}}\right)^2 g^2(z) T^2(k),
\end{equation}
where $\Delta_R$ is the primordial power of the curvature perturbation and is given by
\begin{equation}
  \Delta_R^2=A \left(\frac{k}{k_{\rm CMB}}\right)^{n_s-1}
 \end{equation}
with $k_{\rm CMB}=0.002 {\rm Mpc}^{-1}$, $n_s=0.96$ and
\begin{equation}\label{e.valA}
A=2.41 \times 10^{-9},
\end{equation}
so that it is CMB normalized \cite{WMAP7}.

For a $\Lambda$CDM model, the growing mode of the matter density perturbation is given in terms of a hypergeometric function as \cite{pubook}
\begin{eqnarray}
\!\!\!\!\!\!\!\!\!\!\!\! D_+(t) \!\!&=&\!\! a(t)\, {}_2F_1\left[1,\frac13;\frac{11}{6};-\sinh^{2}\left(\frac{3}{2}\sqrt{\Omega_{\Lambda0}}H_0 t\right)\right],
\end{eqnarray}
so that the growth factor takes the form
\begin{equation}
 \frac{g(z)}{g_\infty} = {}_2F_1\left[1,\frac13;\frac{11}{6};-\frac{\kappa_0^3}{(1+z)^3}\right].
\end{equation}
The transfer function is parameterized as
\begin{equation}
T_0(q) =\frac{L_0(q)}{L_0(q)+C_0(q)q^2}
\end{equation}
with 
\begin{eqnarray}
L_0(q)&=&\ln(2e+1.8q) \\
C_0(q)&=&14.2+\frac{731}{1+62.5q},
\end{eqnarray}
$q$ being defined as
\begin{equation}
q\equiv\frac{k}{13.41 k_{\rm eq}}=\frac{k}{\Omega_{m0} h^2{\rm Mpc}^{-1}}.
\end{equation}


\subsection{Second choice: observation based}\label{subsec4b}

We can also use a form of the power spectrum motivated by observations and, for simplicity, we assume that its form is similar to the one obtained from the APM survey~\cite{APM}
\begin{equation}\label{spectrum}
   P_\delta(k)=\tilde{A} \frac{k}{\left[1+\left(\frac{k}{k_c}\right)^2\right]^{3/2}}
\end{equation}
with $k_c=(1/20) h {\rm Mpc}^{-1}$. The normalization constant $\tilde A$ is related to $\sigma^2_8$
\begin{equation}\label{sigma8}
  \sigma_8^2=\overline{\left(\frac{1}{V_{R_8}} \int \dd^3{\bm x} \, \delta({\bm x})  W_{R_8}(x) \right)^2}
\end{equation}
where $W_{R_8}$ is a top-hat (spherical) window function with radius $R_8=8 h^{-1} {\rm Mpc}$,  
$\sigma^2_8$ is typically of order unity today (see, e.g., Ref.~\cite{FuetAll}) and 
it is equivalent to the variance $\sigma^2_R$ defined in Appendix C.

In the following, we decide to take, as value for $\sigma_8$, the one obtained using as a starting point the theory based power-spectrum described above. 
Namely, in order for the two spectra to match, we compute $\sigma_8^2$ with the inflationary spectrum, which is CMB normalized, and a set of cosmological parameters. We then impose that the value of $\sigma_8^2$ obtained from the APM spectrum is the same, which determines $\tilde A$.  Assuming $h=0.7$ and a standard $\Lambda$CDM model, we get $\tilde A= 4.21\times10^6 \sigma_8^2\, {\rm Mpc}^4$. The details are given in Appendix~C.

\begin{figure}[h!]
\centering
\includegraphics[width=\columnwidth]{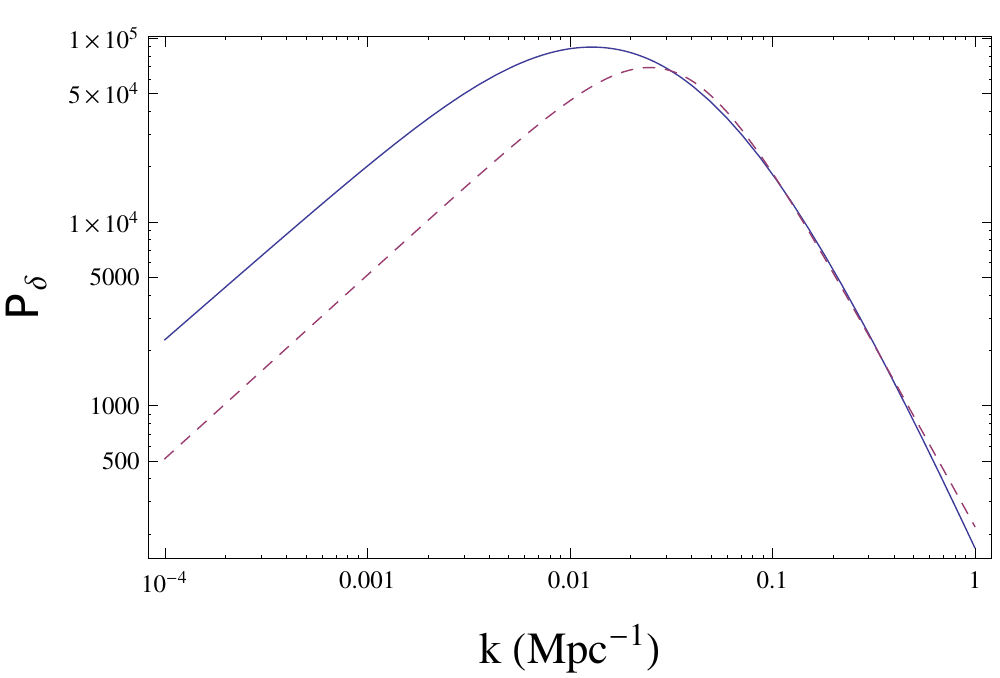}
\caption{Comparison of the matter APM power spectrum ${P}_\delta$ (dashed line) and the $\Lambda$CDM power spectum (solid line)
assuming $(\Omega_{\rm m0},\Omega_{\Lambda0})=(0.27,0.73)$ and a spectral index $n_s=0.96$.}
\label{fig1}
\end{figure}

The power spectrum for $\Psi$, given by Eqs.~(\ref{ConnectionPowerSpectra}) and (\ref{spectrum}), is almost constant in the infrared ($k\ll k_c$) and behaves as $k^{-3}$ in the ultraviolet. It thus has the same behaviour as the spectrum obtained from assuming standard inflation in the case of a spectral index $n_s=1$ in the infrared but differs in the ultraviolet. The two power spectra for the matter density contrast are compared in Fig.~\ref{fig1} for $z=0$. As one can note, they are in extremely good agreement in the range of values of $k$ from which we get the main contribution to the shear (see Eqs.~(\ref{Sigma2_k_general0}--\ref{Sigma2_k_general2_w})) and where the non-linear effects can still be neglected. 

\section{Shear Calculation}\label{V}

We are now in a position to actually compute the shear of the averaged flow $\langle \sigma^2 \rangle_{\cal D}$ using these two power spectra. As already mentioned, the ensemble average of the spatially averaged shear loses reference to the averaging domain ${\cal D}$, so that we can drop the dependence on ${\cal D}$. 

For a $\Lambda$CDM model, Eq.~(\ref{Eq1}) implies that $\Psi\propto D_+/a$ so that
\begin{equation}\label{Psi_k_general}
\Psi_{\bm k}(z)=\Psi_{\bm k}(z=0) \frac{g(z)}{g_0}\,,
\end{equation}
where $g_0=g(z=0)$.
Considering Eq.~(\ref{B_Fourier}) to evaluate $B_k$ 
one gets
\begin{equation}
B_{\bm k}=-(1+z)\frac{1}{a_0 H_0} \Psi_{\bm k}(0) \int_{zi}^{z} dz' \frac{H_0}{H(z')(1+z')}
\frac{g(z')}{g_0}\,.
\end{equation}
The integral has to be performed numerically and the solution can be written as
\begin{equation}
B_{\bm k}=-(1+z)\frac{1}{a_0 H_0} \Psi_{\bm k}(0) I(z, \Omega_{m0}, \Omega_{\Lambda0})
\end{equation}
with
\begin{equation}\label{def_I}
I(z, \Omega_{\rm m0}, \Omega_{\Lambda0})=\int_{z_i}^{z} dz' \frac{H_0}{H(z')(1+z')}
\frac{g(z')}{g_0}
\end{equation}
where $z_i$ is an initial redshift deep in the matter era, the choice of which does not impact the final result.

Going on we find that, starting from Eq.~(\ref{Res_EA}), the ensemble average of the shear is given by 
\begin{eqnarray}
  \overline{\langle \sigma^2 \rangle}&=&\frac{1}{6 \pi^2 a^2} \left(\frac{1}{a_0H_0}\right)^2
  (1+z)^2 I^2(z, \Omega_{\rm m0}, \Omega_{\Lambda0}) \nonumber \\
  & &   \int \dd k k^6  |\Psi_k(0)|^2.
\end{eqnarray}

\subsection{APM power spectrum}

Using the expression of the power spectrum~(\ref{ConnectionPowerSpectra}--\ref{spectrum}), the previous computation can be performed analytically
to obtain
\begin{eqnarray}\label{Sigma2_k_general0}
   \frac{\overline{\langle \sigma^2 \rangle}}{3 H^2} &= &\frac{ \tilde A}{8 \pi^2} 
   \frac{H_0^2}{H^2}\Omega_{m0}^2(1+z)^4
   I^2(z, \Omega_{m0}, \Omega_{\Lambda0}) \nonumber \\
  &&  \int \dd k \frac{k^3}{\left[1+(k/k_c)^2\right]^{3/2}}\,.
\end{eqnarray}
This integral is clearly convergent in the infrared but diverges in the ultraviolet.\footnote{We should note that, on the contrary, the value of the expansion rate $(\dot{a}_{{\cal D}}/a_{{\cal D}})$ given in Eq.~(\ref{e.backreac1}) is not divergent in the ultraviolet. This is a consequence of the fact that the divergent term in the shear cancels with a similar divergence in the genuine kinematical backreaction (which at leading order is $\tilde{Q}_{\cal D}\sim 2 \overline{\langle \sigma^2 \rangle}_{\cal D}$). Therefore, this divergent behavior can imprint on the level of anisotropy of the averaged flow while not affecting the total backreaction on $(\dot{a}_{{\cal D}}/a_{{\cal D}})$ if calculated with the standard Buchert procedure.} It can be regularized in two different ways: either by introducing
an UV cut-off $k_{\rm UV}$, or with the method introduced in Ref.~\cite{CAN} in which $|\Psi_{\bm k}(0)|^2$ is replaced by $W(k R_s)^2 |\Psi_{\bm k}(0)|^2$
with $W(k R_s)=e^{-k^2 R_s^2/2}$. $R_s$ is a smoothing scale (and we will assume that $R_s=k^{-1}_{\rm UV}$).

The result of the integration then takes the form
\begin{eqnarray}\label{Sigma2_k_general1}
   \frac{\overline{\langle \sigma^2 \rangle}}{3 H^2} &= &\frac{\tilde A k_c^4}{4 \pi^2}\Omega_{\rm m0}^2
   f(z) W(k_{\rm UV})
\end{eqnarray}   
where $\frac{\tilde A k_c^4}{4 \pi^2}\Omega_{\rm m0}^2$ is a constant but depends on the 
particular set of cosmological parameters considered, the redshift evolution is characterized by
\begin{eqnarray}\label{def_f}
 f(z)\equiv \frac{(1+z)^4}{E(z)}  I^2(z, \Omega_{\rm m0}, \Omega_{\Lambda0})
\end{eqnarray}
and $W(k_{\rm UV})$ is a function depending on the regularization scheme and the cut-off (but not 
on the cosmological parameters). 
With a simple cut-off 
\begin{equation} \label{shearv1}
 W(k_{\rm UV})= -1+\left(1+\frac{k_{\rm UV}^2}{2 k_c^2}\right)
 \left(1+\frac{k_{\rm UV}^2}{k_c^2}\right)^{-1/2},
\end{equation}
while the regularization based on a Gaussian window function gives
\begin{eqnarray} \label{shearv2}
W(k_{\rm UV})&=& - \frac12 +
\frac{\sqrt{\pi}}{4} \frac{k_{\rm UV}}{k_c} e^{k_c^2/k_{\rm UV}^2} 
 \left(1+2 \frac{k_c^2}{k_{\rm UV}^2}\right)\times\nonumber\\
 &&\qquad\times\left[1-{\rm Erf}\left(\frac{k_c}{k_{\rm UV}}\right)\right]
\end{eqnarray}
where ${\rm Erf}$ is the error function.


\subsection{$\Lambda$CDM power spectrum}

Let us now repeat this calculation with the power spectrum~(\ref{PowerPsiStandard}). The expression of the shear is now given by
\begin{eqnarray}\label{Sigma2_k_general2}
\frac{\overline{\langle \sigma^2 \rangle}}{3 H^2}&=&\frac{2 A}{25}\frac{k_c^4}{a_0^4H_0^4}
\left(\frac{g_0}{ g_\infty} \right)^2  f(z) W(k_{\rm UV})
\end{eqnarray}
where we now have
\begin{eqnarray}\label{Sigma2_k_general2_w}
W(k_{\rm UV}) = \frac{1}{2} \int \frac{\dd k k^3}{k_c^4} \left(\frac{k}{k_{\rm CMB}}\right)^{n_s-1}\!\!\!\!  T^2(k).
\end{eqnarray}
This expression is clearly convergent in the infrared while it diverges in the ultraviolet so that we have to regularize it.

Proceeding as in the previous paragraph, the use of an UV cut-off leads to
\begin{eqnarray}\label{Sigma2_k_general2_with_Cutoff}
W =\frac{1}{2} 
\int_0^{k_{\rm UV}} \frac{\dd k k^3}{k_c^4} \left(\frac{k}{k_{\rm CMB}}\right)^{n_s-1}  T^2(k),
\end{eqnarray}
or
\begin{eqnarray}\label{Sigma2_k_general2_with_window}
W =\frac{1}{2} 
\int_0^{\infty} \frac{\dd k k^3}{k_c^4} \left(\frac{k}{k_{\rm CMB}}\right)^{n_s-1} \!\!\!\!  e^{-(k/k_{\rm UV})^2}T^2(k)\,
\end{eqnarray}
for a window function. In both cases, the integration cannot be performed analytically and we have to rely on a numerical integration.
Note, however, that the result will depend only mildly on the spectral index as long as $n_s \sim 1$.

\begin{figure}[h!]
\centering
\includegraphics[width=\columnwidth]{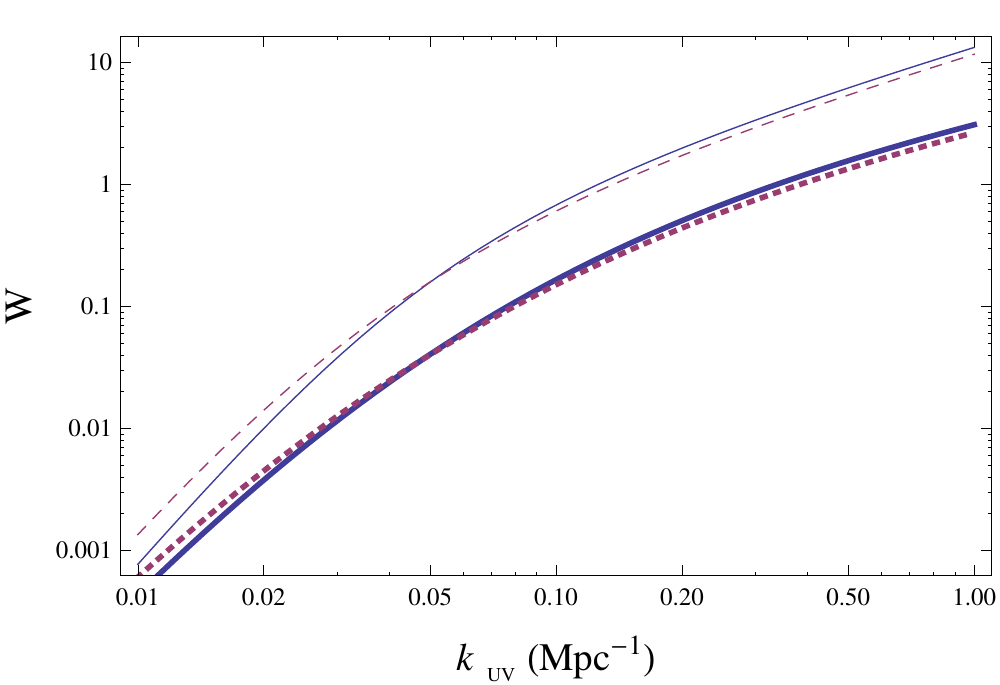}
\caption{Dependence of the window function $W$ on the UV cut-off scales $k_{\rm UV}$, assuming the APM (thin lines) or an inflationary (thick lines) power spectrum. The solid line refers to a sharp cut-off and the dashed line to a Gaussian regularization.}
\label{fig2}
\end{figure}

These two window functions are compared  on Fig. \ref{fig2}, considering both the APM and the inflationary power spectrums.
In both cases, it can be concluded that the dependence on the choice of the window function, 
Eqs.~(\ref{shearv1}) or~(\ref{shearv2}) for the APM power spectrum and 
Eqs.~(\ref{Sigma2_k_general2_with_Cutoff}) or~(\ref{Sigma2_k_general2_with_window}) for the inflationary power spectrum,
 is mild. As long as $k_{\rm UV}$ remains small compared to $1\, {\rm Mpc}^{-1}$, there is almost no difference.
See also Fig.~\ref{fig3} below, where we have depicted, for the APM case, the time evolution of the shear for the two window functions assuming $k_{\rm UV}=0.1\, {\rm Mpc}^{-1}$, $k_{\rm UV}=0.5\, {\rm Mpc}^{-1}$ and $k_{\rm UV}=1\, {\rm Mpc}^{-1}$.

\subsection{Properties of the shear}

The two previous sections show that the choices of the cut-off scale and of the power spectrum affect the overall value of the shear. However, since, as seen from Fig.~\ref{fig1}, the two spectra are very similar in shape, we do not have an important difference between the two choices. It is also important to realize that the redshift dependence of the shear is dictated by the function $f(z)$, which is the same in the two cases.

Let us first discuss the amplitude. The typical amplitude of $\Sigma$ is almost unaffected by the choice of power spectrum. Typically the value of $\Sigma$ computed with the inflationary spectrum will be from 3\% to 9\% smaller than the value computed with the APM spectrum when  $k_{\rm UV}$ changes from $0.1\,{\rm Mpc}^{-1}$ to $1\,{\rm Mpc}^{-1}$ . For that reason, we will now present just one set of results. Fig.~\ref{fig3} depicts the evolution of $\Sigma$ for different values of $k_{\rm UV}$ and 
it can easily be checked that the ratio between the overall amplitudes of two
curves coincides with the ratio of the corresponding two window functions $W(k_{\rm UV})$. 
Typically, the overall amplitude is multiplied by a factor of order $9$ or $19$, respectively, when using $k_{\rm UV}=0.5\,{\rm Mpc}^{-1}$ or $k_{\rm UV}=1\,{\rm Mpc}^{-1}$ instead of $k_{\rm UV}=0.1\,{\rm Mpc}^{-1}$. 

\begin{widetext}.
\begin{figure}[htb]
\centering
 \includegraphics[width=5.5cm]{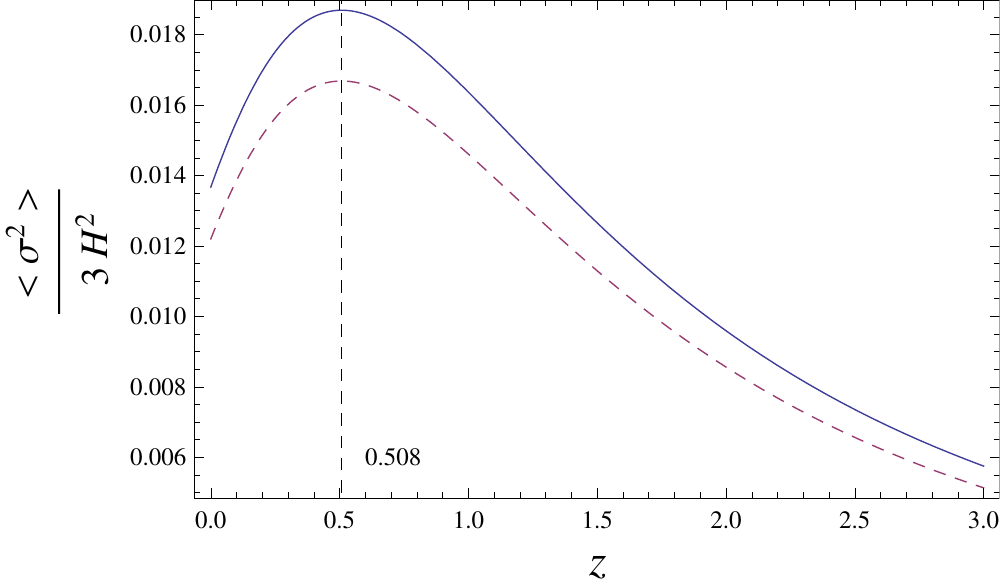}\includegraphics[width=5.5cm]{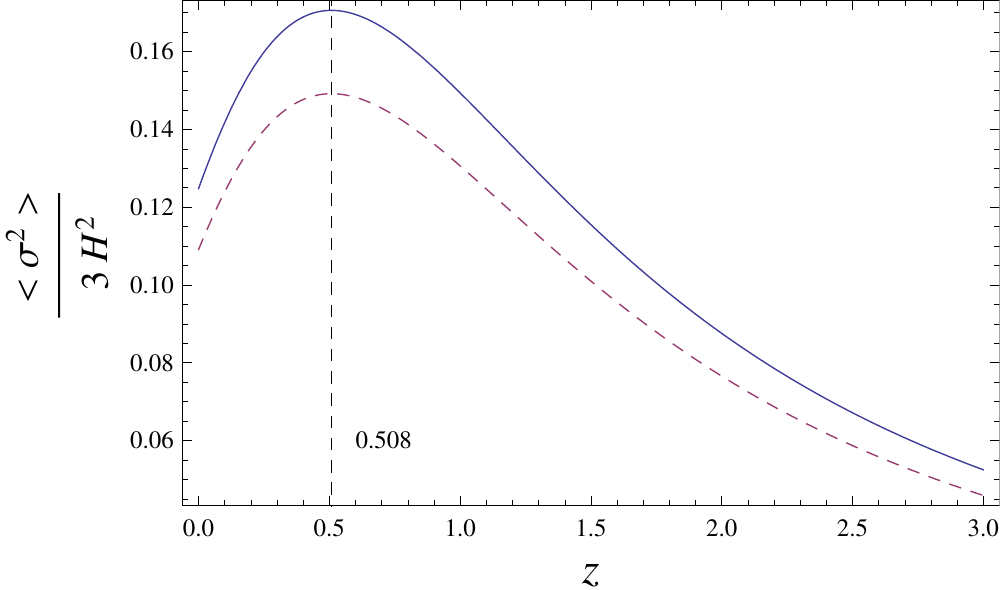} \includegraphics[width=5.5cm]{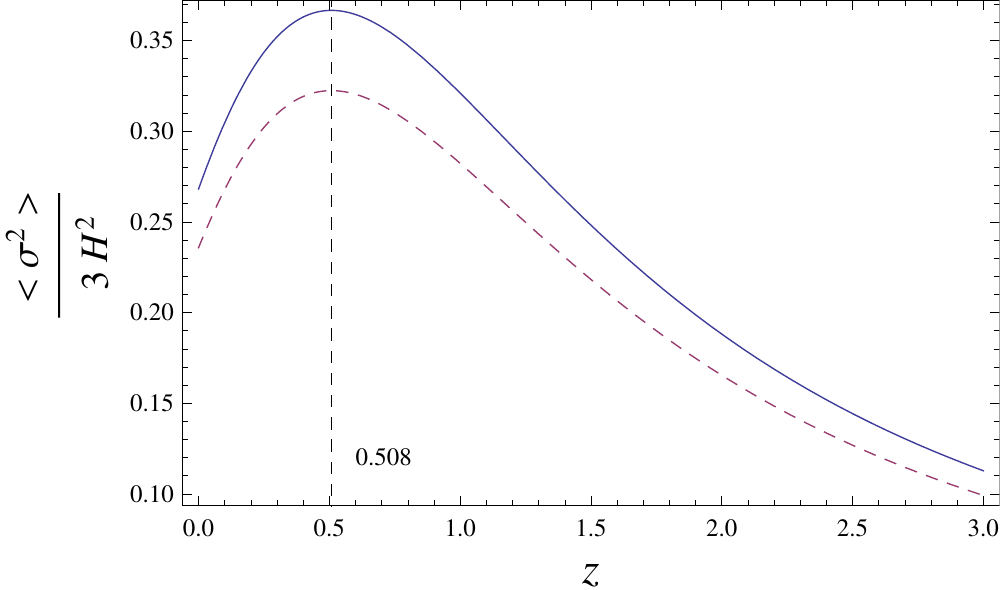} 
\caption{Dependence of the anisotropy $\Sigma$  as a function of the redshift $z$ assuming UV cut-off $k_{\rm UV}=0.1, 0.5, 1\, {\rm Mpc}^{-1}$ (from  left to the right) and assuming the APM power spectrum with $\Omega_{\Lambda0}=0.73$. We plot both the solution (\ref{shearv1}) (solid lines) and the solution (\ref{shearv2}) (dashed lines).}
\label{fig3}
\end{figure}
\end{widetext}

Let us now discuss the time evolution of $\Sigma$, which follows from the form of the function $f(z)$ defined in Eq.~(\ref{def_f}). The evolution of $I(z,\Omega_{\rm m0},\Omega_{\Lambda0})$ at large redshift (typically $z>1$) can be obtained by assuming that $g(z)$ is almost constant and $g(z)\simeq g_\infty$. In this case we have
\begin{equation}
 I(z,\Omega_{\rm m0},\Omega_{\Lambda0}) \simeq \frac{1}{3}\frac{g_\infty}{g_0}\Omega_{\Lambda0}^{-1/2}
 \ln\frac{\sqrt{1+\alpha_0(1+z)^3}-1}{\sqrt{1+\alpha_0(1+z)^3}+1}\nonumber
\end{equation}
with $\alpha_0=\Omega_{\rm m0}/\Omega_{\Lambda0}$. It follows that
\begin{equation}\label{e.asymp}
\frac{\average{\sigma^2}}{3H^2}\simeq \frac{\tilde A k_c^4}{9\pi^2}\left(\frac{g_\infty}{g_0}\right)^2  W(k_{\rm UV})\, \frac{1}{z^2}
\end{equation}
for $z>1$. This is indeed consistent with our numerical results, as depicted on Fig.~\ref{fig3}.

\begin{figure}[htb]
\centering
 \includegraphics[width=8cm]{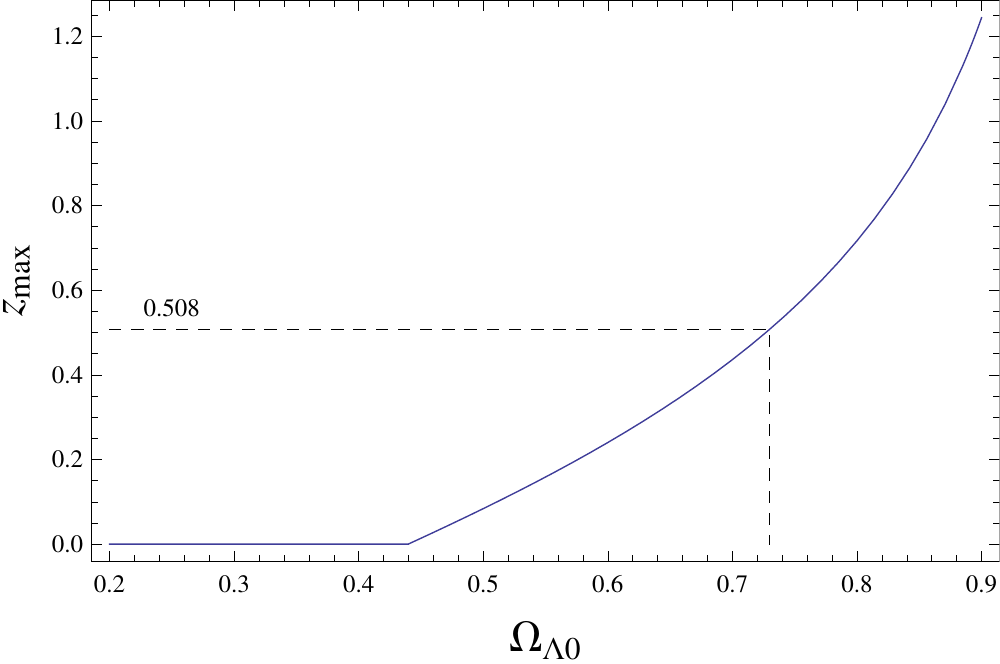}
\caption{Value $z_{\rm max}$ of the position of the peak of $\Sigma$ as a function of $\Omega_{\Lambda0}$. Such a peak exists only if the condition~(\ref{condzmax}) is satisfied, that is for $\Omega_{\Lambda0}>0.44$.}
\label{fig4}
\end{figure}

\begin{figure}[htb]
\centering
 \includegraphics[width=8.5cm]{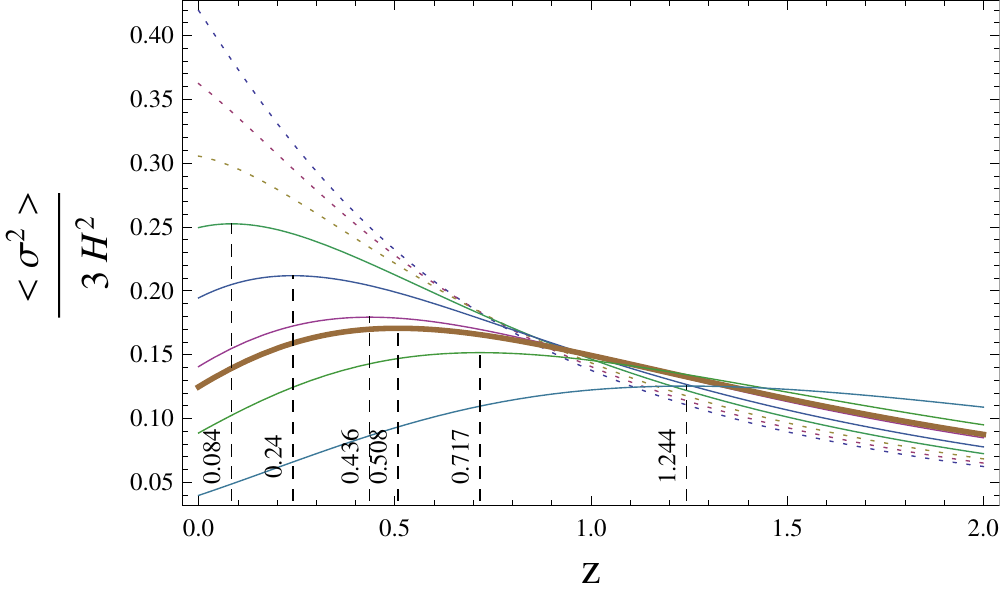}
\caption{Dependence of $\Sigma(z)$ on the value of $\Omega_{\Lambda0}$ for an UV cut-off $k_{\rm UV}=0.5\, {\rm Mpc}^{-1}$. The position of the peak corresponds to the value of $z_{\rm max}$ depicted in Fig.~\ref{fig4}. $\Omega_{\Lambda0}$ ranges from $0.2$ to $0.9$ by steps of $0.1$ (with the addition of the curve correspondent to $\Omega_{\Lambda0}=0.73$). The dashed curves do not satisfy the condition~(\ref{condzmax}) and have no maximum. The thick curve is the one which corresponds to $\Omega_{\Lambda0}=0.73$.}
\label{fig5}
\end{figure}

In $z=0$, the function $f$ can be expanded as $f(z)\simeq f(0) + f'(0) z$ so that the function starts increasing only if $f'(0)>0$, in which case $\Sigma(z)$ will peak at a redshift $z_{\rm max}$. The condition for the existence of such a peak takes the form
\begin{equation}\label{condzmax}
 4-3\Omega_{\rm m0} + \frac{2}{I(0,\Omega_{\rm m0},\Omega_{\Lambda0})}>0
\end{equation}
since from Eq.~(\ref{def_I}), $I'(0,\Omega_{\rm m0},\Omega_{\Lambda0})=1$. This condition cannot be easily translated into a condition on the cosmological parameters, but for a flat $\Lambda$CDM, one gets that there exists a peak if $\Omega_{\Lambda0}>0.44$. Interestingly, the position of the peak depends on the value of the cosmological constant. Fig.~\ref{fig4} depicts the value $z_{\rm max}$ of the position of the peak as a function of $\Omega_{\Lambda0}$. This computation can be compared to the explicit form of $\Sigma(z)$ allowing $\Omega_{\Lambda0}$ to vary between $0.2$ and $0.9$; see Fig.~\ref{fig5}. Note that one needs to compute $\tilde A$ for each value of the cosmological parameters since the relation between $\sigma_8^2$ and the theory based power spectrum depends on the growth rate and $k_{\rm eq}$. 
The amplitude of the peak decreases with $\Omega_{\Lambda0}$ which can be traced back to (1) the fact that the growth rate is smaller for larger $\Lambda$ and (2) the fact that $k_{\rm eq}$ decreases with $\Omega_{\Lambda0}$. 
We will have a longer radiation era with more modes entering the Hubble radius
during this era. As a consequence, they will be significantly damped with respect to those 
which become sub-Hubble during the matter era (see, for example, Ref.~\cite{CU}).

\begin{figure}[htb]
\centering
 \includegraphics[width=8cm]{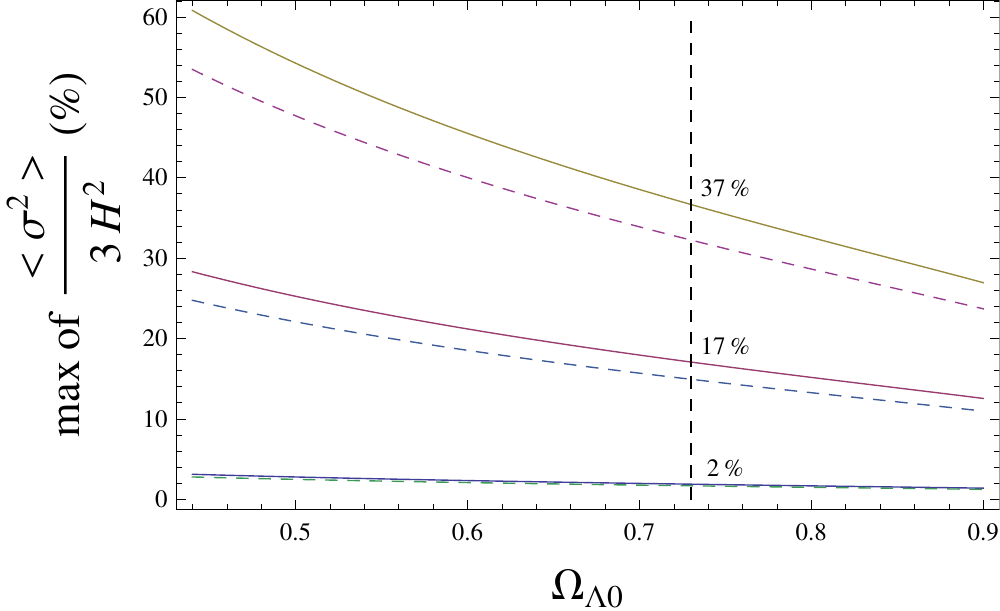}
\caption{Amplitude of the peak, $\Sigma(z_{\rm max})$ as a function of $\Omega_{\Lambda0}$. From bottom to top $k_{\rm UV}=0.1, 0.5, 1\, {\rm Mpc}^{-1}$ and we have used the two regularization schemes (solution (\ref{shearv1}) (solid lines) and solution (\ref{shearv2}) (dashed lines)). It decreases with $\Omega_{\Lambda0}$.}
\label{fig6}
\end{figure}

The behavior of $\Sigma(z)$ has to be compared with the behavior of a homogeneous shear, e.g., in a Bianchi~I universe, for which it scales as $\sigma^2\propto (1+z)^6$, so that in the matter era, it decays as $\Sigma_{\rm Bianchi}\sim z^3$~\cite{ppu} while the spatially average anisotropy induced by the large scale structure grows as $\Sigma\sim z^{-2}$ to eventually peak and then slowly decay as $\Sigma\sim z$. Its typical amplitude is set by the value of the cosmological parameters and the normalization of the matter power spectrum, contrary to a Bianchi universe for which it is a pure initial condition.


\section{Discussion}\label{VI}

This article relies on the simple remark that generically a spatially averaged flow has no reason to be isotropic. We have thus focused our analysis of the shear in order to quantify the expected deviation from isotropy that arises from the backreaction of large scale structure. 

For that purpose, we have used the averaging procedure designed in Ref.~\cite{Buchert} but rewritten in a way that makes explicit the anisotropy of the averaged flow. Under this new form the kinematical backreaction term arises only from inhomogeneities and, e.g., the spatial average of a Bianchi dynamics remains unchanged, so that it is physically more sound (in particular, one can think of comparing the backreaction that arises from similar inhomogeneities with respect to different background homogeneous flows). In order to evaluate the order of magnitude of the scalar anisotropy $\Sigma$, it is sufficient to work  at first order in perturbation theory.\\

Concerning the questions that were raised in the introduction, we can give the following answers. 

First, $\overline{\langle \sigma^2 \rangle}$ diverges in the ultraviolet and it must be regularized by the introduction of a UV cut-off $k_{\rm UV}$; see Eqs.~(\ref{shearv1}) and~(\ref{shearv2}) or Eqs.~(\ref{Sigma2_k_general2_with_Cutoff}) and~(\ref{Sigma2_k_general2_with_window}). Such UV divergence has a large impact on the final result and the same problem of UV divergent terms is also found in other approaches (e.g.  the one based on the all-sky average (monopole) of the redshift-distance relation, obtained from the Kristan and Sachs approach \cite{KS}, as clearly shown in Ref.~\cite{CU}).\footnote{However, note that the use of a well-defined  procedure to average over the light-cone~\cite{GMNV}, applied to the  perturbative expansion up to the second order of the exact redshift-distance relation, gives a well-defined result with no ultraviolet divergences \cite{BGMNV}.} The question regarding this cut-off is central in the backreaction debate. Indeed, since we have used linear perturbation theory, it is probably not safe to extrapolate at scales larger than $k\sim1 \,{\rm Mpc}^{-1}$ without taking into account the non-linear effects which turn on beyond such scales. This divergence is particulary problematic for the shear,  since this is related to an anisotropy of the Hubble flow and can thus, in principle, be observed. We see that its amplitude can shift from 2\% to  37\% when $k_{\rm UV}$ varies from $0.1$ to $1\,{\rm Mpc}^{-1}$ for the standard $\Lambda$CDM model 
(see Fig. 3), and it varies with the value of the cosmological constant (see Fig. 6).

Since the amplitude of some backreaction effects can also depend on $k_{\rm UV}$ (see Ref.~\cite{CU}), a constraint of the scalar shear can be translated into a constraint on the  backreaction, hence possibly resolving the debate of the ability for these effects to explain the late time acceleration of the universe. The average Friedmann equation~(\ref{e.backreac1bis}) can be rewritten as $1=\Omega_{\rm m} + \Omega_\Lambda + \Omega_\sigma + \Omega_K + \tilde\Omega_Q$. It is well-known (see, for example, Refs.~\cite{LS,CAN}) that $\Omega_Q=\Omega_\sigma + \tilde\Omega_Q$, which usually characterizes the backreaction, is not UV divergent, while both $\Omega_\sigma$ and $\tilde\Omega_Q$ are. This arises from the fact that the divergence does not appear in $\average{\Theta}$ while it does in both $\average{\sigma}$ and $\average{\Theta^2}$. This shows trivially that $\Omega_\sigma\sim- \tilde\Omega_Q$ for large $k_{\rm UV}$ so that the amplitude of the anisotropy and kinematical backreaction are related in this regime.

Let us stress that there is no total agreement about the physical meaning  of the Buchert procedure and its connection to observable quantities. For example, quantities  connected to spatial averages are not, strictly speaking, observable because we only have observational access to our past light-cone (see, for example, Ref.~\cite{CU}). On the other hand, if we take seriously these equations we have an important observational impact via the averaged shear.  

It is then important to realize that the anisotropy arising from the averaged flow, as a specific signature of backreaction, may allow us to discriminate with dark energy models~\cite{jpude} (unless dark energy has an anisotropic stress, as e.g. in Ref.~\cite{anisoDE}) or modification of general relativity. We argue that this is a key quantity to discriminate these models as a possible explanation of the late time acceleration of the cosmic expansion. We have shown that its amplitude is fixed by the cosmological parameters and the matter power spectrum, and not as an initial condition as e.g. for Bianchi universes. Its time evolution is also different from the homogeneous shear of a Bianchi universe since it increases with time to eventually reach a maximum. In an effective way, this can indeed be rephrased as a dark energy anisotropic stress, the time evolution of which is determined by the growth of  the structures.

We have shown that for a standard $\Lambda$CDM background dynamics, $\Sigma$ peaks between 2 and 37\% at a redshift of order 0.5. Any bound on the anisotropy of the Hubble flow can probably set constraints on backreaction effects. In particular, it shows explicitly that a cut-off scales larger than or of the order of  $1~{\rm Mpc}^{-1}$ induces too strong an anisotropy. Indeed, it does not make sense to derive the value of a cut-off from observation, but it emphasizes, again, the necessity to design a proper regularization scheme for this mechanism to be fully predictive. An anisotropy with that amplitude can potentially be observable, in particular with weak lensing experiment such as Euclid~\cite{euclid} via the $B$-modes~\cite{pup}.

From a technical point of view, we have used two matter power spectra. One is more realistic while the second has a more simple form. Interestingly, we have shown that they give similar results as long as $k_{\rm UV}$ remains small. This is thus a excellent tool that allows us to make analytical estimates and, for instance, to derive scalings and order of magnitude without numerical integrations of the spectrum.

Our analysis, while restricted to simple power spectra and linear perturbation theory, points toward a specific signature of backreaction. This may offer the possibility to constrain this class of explanations for the dark energy.

\vskip0.5cm
{\bf Acknowledgements:} We thank Chris Clarkson, Obinnah Umeh and Carlo Schimd for discussions and the Action Sp\'ecifique GRAM for its financial support. We also thank Thomas Buchert for his comments and insights.

\appendix

\section{Linear peturbation theory}

We summarize here the main result of linear perturbation theory around a spatially Euclidean FL spacetime. We use the notation of Ref.~\cite{pubook}. the standard scalar-vector-tensor  decomposition of the metric component takes the form
\begin{eqnarray} \label{General_metric}
\!\!\! \!\!\! \delta^{(1)}  g_{00} &=& -2 a^2 A \,,\nonumber\\
 \!\!\!\!\!\!  \delta^{(1)}  g_{i0}&=& a^2 B_i=a^2 (\partial_i B+\bar{B}_i) \,,\nonumber\\
\!\!\! \!\!\!  \delta^{(1)}g_{ij} &=& a^2 \left[ \delta_{ij} 2 C +2 \partial_i \partial_j E +
2 \partial_{(i}\bar{E}_{j)} +2 \bar{E}_{i j} \right]\,,\nonumber
\end{eqnarray}
with 4 scalar degrees of freedom ($A$, $B$, $C$ and $E$), 2 transverse vectors ($\bar{B}_i$ 
and $\bar{E}_i$ with $\partial^i \bar{B}_i=0$,  $\partial^i \bar{E}_i=0$) with 4 degrees of freedom, and a traceless and transverse tensor ($\bar{E}_{ij}$ with
$ \partial^i \bar{E}_{ij}=0=\bar{E}_i^i$) with 2 degrees of freedom. One can then define 6 gauge invariant degrees of freedom 
usually defined as
\begin{eqnarray}
\Psi &\equiv& -C-\mathcal{H}(B-E')\,, \\
\Phi &\equiv& A+\mathcal{H}(B-E')+(B-E')'\,, \\
\bar{\Phi}^i &\equiv& \bar{E}^{i'}-\bar{B}^{i}\,, \\
\bar{E}^{i j}&&
\end{eqnarray}
where the prime denotes the derivative with respect to conformal time and ${\cal H}=a'/a$.

Considering a matter sector described by a perfect fluid with stress-energy tensor
\begin{equation}
T_{\mu \nu}=(\rho+P)u_\mu u_\nu+P g_{\mu\nu}\,,
\end{equation}
where the density and pressure can be splitted as $\rho({\bm x}, \eta)=\rho(\eta)+\delta \rho ({\bm x}, \eta)$ and
$P({\bm x}, \eta)=P(\eta)+\delta P ({\bm x}, \eta)$, and the velocity of the comoving observers is decomposed as
$u^\mu=\bar{u}^\mu+\delta u^\mu$ with $u_\mu u^\mu=-1$. It follows that
\begin{equation}
u^\mu=a^{-1}(1-A, v^i)\,\,\,\,,\,\,\,u_\mu=a (-1-A, v_i+B_i)\,
\end{equation}
and we decompose $v_i$ into scalar and a vector component according to
\begin{equation}
  v_i=\partial_i v+\bar{v}_i\,.
\end{equation}
Some of the gauge invariant variables associated to the matter sector are then
given by
\begin{eqnarray}
\delta^C&=&\delta+\frac{\rho'}{\rho}(v+B) \\
V&=&v+E' \\
\bar{V}_i&=&\bar{v}_i+\bar{B}_i.
\end{eqnarray}

The scalar shear, given in Eq.~(\ref{sigma2}), is by construction a second order quantity so that
\begin{equation} \label{shear0and1}
 (\sigma^2)^{(0)}= (\sigma^2)^{(1)}=0.
\end{equation}
Thus, at the lowest order and for a general metric, we have
\begin{eqnarray}
(\sigma^2)^{(2)}&=&\frac{1}{2 a^2 {S'}^{2}}\left[\delta S_{,i j}
\delta S^{,i j}-\frac{1}{3}(\nabla^2 \delta S)^2 \right]
\nonumber \\ & &
+\frac{1}{2 a^2}\left[B_{i, j}
B^{i, j}-\frac{1}{3}(\partial^i B_i)^2 \right] \nonumber\\
& & +\frac{1}{a^2 S'}\left[\delta S_{,i j}
B^{i, j}-\frac{1}{3}(\nabla^2 \delta S) (\partial^i B_i) \right]
\nonumber \\ & &
-\frac{1}{a^2 {S'}} \delta S_{,i j} {\tilde{h}}^{',i
j}-\frac{1}{a^2}  B_{i, j} {\tilde{h}}^{',i j}
\nonumber \\ & &
+\frac{1}{2 a^2}
{\tilde{h}}_{',i j} {\tilde{h}}^{',i j}\,,
\label{sigma2order2general}
\end{eqnarray}
where  $\tilde{h}_{,i j}=\partial_i \partial_j E-\frac{1}{3}\delta_{ij} \nabla^2 E+\partial_i \bar{E}_j+\partial_j \bar{E}_i+\bar{E}_{i j}$, and we use the notation $X_{,i}\equiv \partial_i X$ for any field $X$. 

It is clear that first order perturbation theory is sufficient to obtain the general expression for the shear up to second order, since second order perturbations will contribute only to third or fourth order to $\sigma^2$.

Let us stress that if we neglect vector and tensor perturbations at first order (considering their contribution negligible with respect to the scalar one) then
there will be no vector or tensor contributions in $(\sigma^2)^{(2)}$.  In fact, genuine second order vector and tensor contributions, that are always sourced by first order scalar perturbations, are also not present in $(\sigma^2)^{(2)}$ as we have seen. We can note that this is also a consequence of a general property valid for any scalar field at any order in perturbation theory. Namely, at a given order $n$ the vectors $\bar{B}_i^{(n)}$ and $\bar{E}_i^{(n)}$, and the tensor $h_{i j}^{(n)}$ can appear in a scalar quantity only as $\partial^i \bar{B}_i^{(n)}$, $\partial^i \bar{E}_i^{(n)}$,  $\partial^i \partial^j h_{i j}^{(n)}$ or $h_{i}^{j\,(n)}$, but since the vectors and the tensors are, respectively, transverse, and traceless and transverse  to all order, these terms are always identically zero. 


\section{Shear average}

Following Refs.~\cite{GMV1, GMV2}, the spatial average prescription described in Eq.~(\ref{averagepb}) can be generalized to take a manifestly gauge invariant form, which allows the use of different gauges independently of the choice of the spacelike hypersurface on which the average is performed.
For instance, the average of a scalar quantity $A(t,{\bm x})$ on a hypersurface $S(t,{\bm x})=S_0$ can be obtained from the four-dimensional integral
\begin{eqnarray}\label{4dimInt}
  I(A,\Omega) &=& \int_{\Omega(t, {\bm x})} \dd^{4} x  \sqrt{-g(t,{\bm x})} \,A(t,{\bm x}) \nonumber \\
  &\equiv& \int \dd^{4} x  \sqrt{-g(t,{\bm x})} \,A(t,{\bm x})W_\Omega(t,{\bm x})\,.
\end{eqnarray}
Here $g$ is the determinant of the 4-metric $g_{\mu\nu}$ and the window function is given by
\begin{equation}\label{WindowFunction}
  W_\Omega(t,{\bm x})= n^{\mu} \nabla_\mu \theta[S(t,{\bm x})-S_0] \tilde{W}_\Omega(t,{\bm x}),
\end{equation}
where $\tilde{W}_\Omega(t,{\bm x})$ defines the 3-dimensional domain ${\cal D}$ 
inside the 3-dimensional hypersurface $S(t,{\bm x})=S_0$.
Then, the average is simply given by~\cite{GMV1, GMV2}
\begin{widetext}
\begin{equation}
\langle D(x) \rangle_{\{S_0,r_0\}}= 
{\int_{{\Sigma}_{S_0}} \dd^3{\bm x}  \sqrt{|\overline{\gamma}(t_0, {\bm x})|} 
\,~ \overline{D}(t_0, {\bm x}) \,\tilde{W}_\Omega[h(t_0, {\bm x}), {\bm x})]
\over \int_{{\Sigma}_{S_0}}  \dd^3{\bm x}  \sqrt{|\overline{\gamma}(t_0, {\bm x})|} \,  
\tilde{W}_\Omega[h(t_0, {\bm x}), {\bm x})]},
\label{36}
\end{equation}
\end{widetext}
where $\gamma$ is the determinant of the 3-metric $g_{i j}$ and the bar indicates that the 
quantities are evaluated in a new
bar coordinate system in which the scalar $S$ is homogeneous (i.e. $S(t,{\bm x})=S^{(0)}(t)$).
This corresponds to a redefinition of the time coordinate, $t\rightarrow\bar{t}$ with 
$t=h(\bar{t}, {\bm x})$.
Furthermore, $t_0$ is the time $\bar{t}$ at which $S^{(0)}(\bar{t})$ takes
the constant values $S_0$. The suffix $\Sigma_{S_0}$ on the integral recalls that we are averaging a scalar quantity $A$ on a section
of the three-dimensional hypersurface $\Sigma_{S_0}$,  where the given scalar field  $S(t,{\bm x})$  takes the constant values $S_0$.

On the other hand, for the aim of the paper, the simpler definition given by Eq.~(\ref{averagepb})
is sufficient to average the scalar shear without running into problems related to the gauge choice.
In fact, a double simplification appears as a consequence of 
the fact that the scalar shear is non-vanishing only from the second order on (see Eqs.(\ref{shear0and1MT}) and (\ref{sigma2order2generalMT})).
First, performing the computations only up to second order, we can take 
$\sqrt{|\overline{\gamma}|}= \sqrt{|\overline{\gamma}^{(0)}|}=
a(t)^3$, and the particular form of the perturbations of the spatial metric is irrelevant.
Second, the shear will be gauge invariant up to this second order, {\em i.e.} $\left(\bar{\sigma}^2\right)^{(2)}=\left(\sigma^2\right)^{(2)}$. Therefore there is no
difference between the particular bar gauge and any other gauge for such a quantity. 
As a final result, we obtain that, up to second order in perturbation theory,
the average of the shear over a 3-dimensional domain ${\cal D}$
can be simply written as
\begin{equation}
\langle \sigma^2 \rangle_{\cal D}=
\frac{1}{V_{\cal D}} \int_{\cal D} \dd^3 {\bm x} (\sigma^2)^{(2)}(t_0,{\bm x})\,.
\end{equation}
Considering now the gauge introduced in Eq.~(\ref{metric}) of \S~\ref{III.b}, using Eq.~(\ref{sigma2order2generalMT}) and neglecting tensor and vector contributions, we end up with
\begin{equation}\label{sigma_starting_eq}
\langle \sigma^2 \rangle_{\cal D}  =
\frac{1}{V_{\cal D}} \int_{\cal D}  \dd^3 {\bm x} \frac{1}{2 a^2}
 \left[B_{,i j}
B^{,i j}-\frac{1}{3}(\nabla^2 B)^2\right].
\end{equation}

Let us now consider the Fourier expansion of the first order
perturbation $B$ as
\begin{equation}
B(\vec{x}, \eta) = \int\frac{ \dd^3{\bm k}}{(2 \pi)^{3/2}}  \, e^{i
{\bm k}\cdot {\bm x}} B_{\bm{k}}(\eta) \,.
\end{equation}
The integral becomes
\begin{eqnarray}
 & &\langle \sigma^2 \rangle_{\cal D}=\frac{1}{V_{\cal D}}
 \int_{\cal D}  \dd^3 {\bm x} \int \frac{\dd^3 {\bm k}'}{(2 \pi)^3} \dd^3 {\bm k}'' 
 \left\{\left(k'_i k'_j k^{\prime\prime i}k^{\prime\prime j}
 \right. \right.\nonumber \\
 & &  \left.\left. -\frac{1}{3}k^{\prime2}k^{\prime\prime 2}\right)
 B_{{\bm k}'}B_{{\bm k}''}\,
e^{i({\bm k}'+{\bm k}'')\cdot {\bm x}} \right\}.
\end{eqnarray}
If, as predicted from inflation, the cosmological perturbations are related to the 
primordial perturbations which enjoy Gaussian statistics with zero mean~\cite{pubook}, the average can be computed
by taking an ensemble average over many domains~\cite{Ruth}. We denote this additional average by an over-bar.
As a consequence, if $B$ is statistically homogeneous then 
\begin{equation}
\overline{B({\bm k}) B({\bm k}')}=|B_k| \, \delta^{(3)}({\bm k}+{\bm k}').
\end{equation}
The ensemble average of $\langle \sigma^2 \rangle_{\cal D}$ is thus
given by
\begin{eqnarray}
\overline{\langle \sigma^2 \rangle}_{\cal D} &=&
\frac{1}{V_{\cal D}} \int_{\cal D} \dd^3 {\bm x}  \frac{1}{2 a^2}\frac{1}{(2 \pi)^{3}}
\int \dd^3{\bm k}' \frac{2}{3}|k'|^4 |B_{k'}|^2 \nonumber \\
&=& 
\frac{1}{3 a^2}\frac{1}{(2 \pi)^{3}}
\int d^3 k' |k'|^4 |B_{k'}|^2 
\label{B9}
\end{eqnarray}
and it will be independent from the domain of integration ${\cal D}$, whatever is the window function which defines such domain of integration. 
This is a general property for the ensemble average of the spatial average of a second order contribution.
However, it is worth underlining that the result (\ref{B9}) holds only because the higher contributions
arising from the ensemble average, and involving also the induced metric (which is also 
present in the definition of $V_{\cal D}$), will be at least fourth order in perturbation theory.


\section{Normalization of the matter power spectra}

\subsection{Definition of $\sigma_R^2$}

For any density field $\delta$, that can be decomposed in Fourier modes as
\begin{equation}\label{e.TF}
 \delta({\bm x},t) = \int \frac{\dd^3{\bm k}}{(2\pi)^{3/2}}\delta_{\bm k}\hbox{e}^{i {\bm k}.{\bm x}}
\end{equation}
and any window function $W_R(\bm{x})$ of typical width $R$, the smoothed density field is given by
\begin{equation}
 \delta_R({\bm x},t) = \int \dd^3{\bm y}\delta({\bm y}, t)\, W_R(|{\bm x}-{\bm y}|).
\end{equation}
Decomposing the window function in Fourier modes $W_{\bm k}$, as in Eq.~(\ref{e.TF}), it is easily concluded that the Fourier components of the smoothed density field are
\begin{equation}
 \delta_R({\bm k},t) = (2 \pi)^{3/2} \delta_{\bm k}W_{\bm k}.
\end{equation}
It is then clear that the variance $\sigma_R^2\equiv\overline{\delta^2_R({\bm x},t)}$ can be expressed as
\begin{equation}
 \sigma_R^2= \int\frac{\dd k}{k} \frac{k^3}{2\pi^2}P_\delta(k)W_k^2 = \int\frac{\dd k}{k} {\cal P}_\delta(k)W_k^2.
\end{equation}

For a top-hat filter of width $R$ we have that
\begin{equation}
 W_k=\frac{1}{(2 \pi)^{3/2}}\,\frac{3j_1(kR)}{kR}
\end{equation}
where $j_1$ is a spherical Bessel function of order 1 and is explicitly given by
\begin{equation}
 j_1(x) =\frac{\sin x - x\cos x}{x^2}.
\end{equation}
It follows that
\begin{equation}
 \sigma_R^2  = \int\frac{\dd k}{k} {\cal P}_\delta(k)\left[\frac{3j_1(kR)}{kR}\right]^2.
\end{equation}

\subsection{Determination of $\tilde A$}

\begin{figure}[h!]
\centering
\includegraphics[width=\columnwidth]{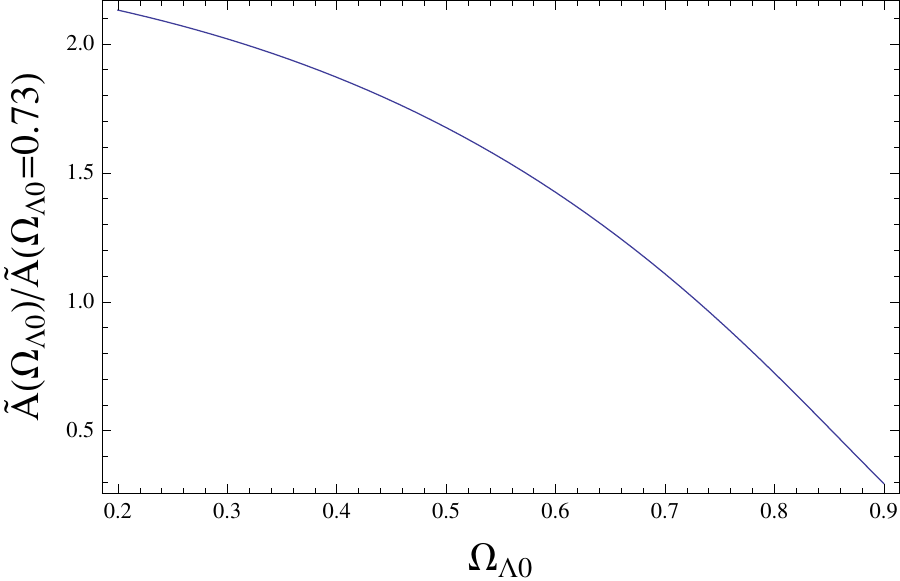}
\caption{Dependence of $\tilde A$ on the cosmological parameters, normalized to its value for $\Omega_{\Lambda0}=0.73$.}
\label{figAtilde}
\end{figure}

The power spectrum determined by inflation is CMB normalized, which fixes the value of $A$ (see Eq.~(\ref{e.valA})). The power spectrum of the gravitational potential is given by Eq.~(\ref{PowerPsiStandard}) which implies that the matter power spectrum at redshift $z=0$ is given, after use of the Poisson equation~(\ref{Eq1}) by
\begin{equation}
 {\cal P}_\delta(k)=\frac{4}{9} \frac{k^4}{\Omega_{\rm m0}^2 a_0^4 H_0^4} {\cal P}_\Psi(k).
\end{equation}
It follows that the value of $\sigma_8^2$ determined from this spectrum is, using the definitions of Section~\ref{subsec4a}, given by
\begin{eqnarray}
\!\!\!\!\!\!\!\!\!\!\!\! \sigma^2_8\equiv \sigma_{R_8}^2 & =& \left(\frac{2}{5}\frac{g_0}{g_\infty}\right)^2\frac{A}{\Omega_{\rm m0}^2 a_0^4} \times\nonumber\\ & &
\!\!\!\!\!\! \!\!\!\!\!\!\!\!\!\!\!\!\! \int\frac{\dd k}{k} \frac{k^4}{H_0^4}\left(\frac{k}{k_{\rm CMB}}\right)^{n_s-1}\!\!\!\!\!\! T^2(k) \left(\frac{3j_1(k R_8)}{k R_8}\right)^2,
\end{eqnarray}
with $R_8=8h^{-1}{\rm Mpc}$. This value of $\sigma_8^2$ depends on the cosmological parameters, in particular via the ratio $g_0/g_\infty$ and $k_{\rm eq}$,
and on $n_s$.

In order to determine $\tilde A$, we impose that the two power spectra give the same $\sigma_8^2$. With the notations of Section~\ref{subsec4b}, we have that
\begin{equation}
 \sigma_8^2 = \tilde A k_c^4 \int\frac{\dd u}{2\pi^2} \frac{u^3}{(1+u^2)^{3/2}}\left[\frac{3j_1(\alpha u)}{\alpha u}\right]^2
\end{equation}
with $\alpha = R_8 k_c=8h^{-1}{\rm Mpc}\times h/20\,{\rm Mpc}^{-1}=2/5$. This latter integral evaluates to $\sim 0.158$ so that
$\tilde A \, k_c^4\simeq 6.32 \, \sigma_8^2$. We thus use
\begin{eqnarray}
 \tilde A &=& 6.32\, k_c^{-4} \sigma_8^2(A,\Omega_{\rm m0},\Omega_{\Lambda0},n_s)\nonumber\\
             &\simeq & 1.01\times10^6 h^{-4} \sigma_8^2(A,\Omega_{\rm m0},\Omega_{\Lambda0},n_s)\,{\rm Mpc}^4.
\end{eqnarray}
Such a choice allows us to tune the APM power spectrum to match the inflationary power spectrum with the drawback that it is not CMB normalized (see Fig.~\ref{fig1}). As explained in the text, this is not too much of a problem since we only use it to derive analytical orders of magnitude and since the computations can be performed numerically with the inflationary power spectrum. The scaling of $\tilde A$ with the cosmological parameters is depicted in Fig.~\ref{figAtilde}.


\end{document}